\documentclass[conference]{IEEEtran}
\IEEEoverridecommandlockouts
\usepackage{cite}
\usepackage{amsmath,amssymb,amsfonts}
\usepackage{algorithmic}
\usepackage{graphicx}
\usepackage{textcomp}
\usepackage{xcolor}
\usepackage{lipsum}
\usepackage{mathtools}
\usepackage{cuted}
\usepackage{amsmath}
\usepackage{bm}
\usepackage{amssymb} 
\usepackage{epsfig} 
\usepackage{subfigure}
\usepackage{textcomp}
\usepackage[utf8]{inputenc}
\usepackage[english]{babel}
\usepackage{lipsum}
\usepackage{mathtools}
\usepackage{cuted}
\usepackage{amsthm}

\usepackage{mathtools}
\usepackage{siunitx}
\DeclarePairedDelimiterXPP\BigOSI[2]%
  {\mathcal{O}}{(}{)}{}%
  {\SI{#1}{#2}}

\begin{document}

\title{{Minimum Phase Linear Antenna Array Design}}

% Note to editor: all contact details and declarations added here
\author{J.C. Olivier and E. Barnard % 
\thanks{ J.C. Olivier is with the School of Engineering, University of Tasmania, Hobart, Australia.  Email:  jc.olivier@utas.edu.au}% 
\thanks{E. Barnard is with The Faculty of Engineering, North-West University, Potchefstroom, South Africa.  Email: etienne.barnard@gmail.com  }
}

\maketitle

\begin{abstract}
The paper considers the design of minimum phase discrete linear arrays. The paper introduces recent advances for  the design of minimum phase Finite Impulse Response filters, as applied to the design of minimum phase linear arrays. The minimum phase linear array is demonstrated to require the least number of elements of all linear arrays that are able to achieve a given  magnitude pattern specification. Three example designs are presented and compared to results from the literature. 
\end{abstract}


% Note to reviewers:  The introduction has remained unchanged. 
\section{Introduction}

The theory and practice of linear antenna array design  are mature and well understood \cite{dolph,hanson}. The array elements are spaced uniformly at $d$ meters, and the element excitation weights are denoted as column vector $\mathbf c$ in this paper.  With the angle relative to the array broadside denoted as $\theta$,  the far field radiation pattern intensity $C$ is given by 
\begin{equation}
    C(\theta) = \sum_{k=0}^{N-1} c(k) e^{j 2 \pi k \frac{d}{\lambda} \sin \theta}.
\end{equation}
 The design objective is to obtain the vector of element weights $\mathbf c$ so that a given specification for the radiation pattern $C(\theta)$ is met (or exceeded). 

With a parameter $u$ defined as 
\begin{equation}
    u = 2 \pi \frac{d}{\lambda} \sin \theta,
\end{equation}
the array synthesis problem is mathematically identical to that of Finite Impulse Response (FIR) filter design.  The parameter $u$ denotes the frequency domain (spectral) representation of the filter with a domain $-\pi \le u \le \pi$. 

\begin{figure} []
\centering
  \includegraphics[width=0.4\textwidth]{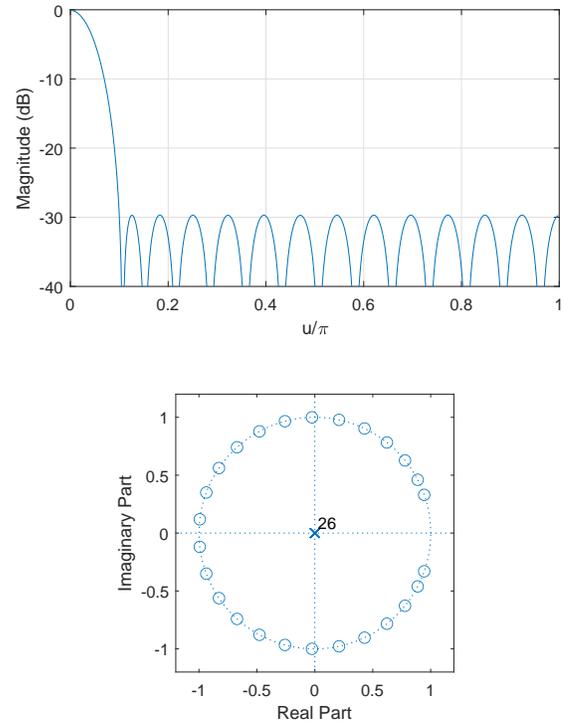}
\caption{A narrow pencil beam $27$ element array  designed with the PM FIR algorithm.  The beamwidth was set to $u = 0.1 \pi$ and the sidelobe maximum was specified as $-30$ dB. As shown in the bottom figure, all zeros are located on the unit circle.}
\label{pencil_FIR}
\end{figure}

For FIR filter design, an optimal Chebyshev approximation was proposed during the  1970's \cite{parks}, and the so-called Parks-McClellan (PM) design is available in MATLAB via the \emph{firpm} function\footnote{The PM algorithm is based on the Remez  approximation theorem \cite{remez}.}.  The PM design produces an FIR filter with linear phase, and hence symmetric real discrete time-domain filter taps.  If the passband cutoff frequency of the lowpass FIR filter is set to zero, then the PM design yields the well known pencil beam  Chebyshev approximation as shown in Figure \ref{pencil_FIR}. 
 The beam direction can be translated to any finite value for $u$ (i.e. $\theta$) by deploying a constant phase shift across the array, yielding complex array element excitation values. 
 
A linear array with the phase not specified is also possible, and under those conditions only the magnitude of the radiation pattern $|C(\theta)|$ is specified \cite{orchard}. This is often the case  if the linear array is expected to have a \emph{flat top}, a case that has received renewed attention in recent literature \cite{flattop1,flattop2,flattop3}.  When the phase is not specified, a so-called \emph{minimum phase linear array} can be designed instead.  Recent advances in the design of an optimal minimum phase FIR filter have been reported in  \cite{antonio,olivier}; these can be employed to design and evaluate the performance of minimum phase linear arrays -- the objective of this paper.  

A minimum phase linear array has a Z-plane representation where all the zeros are inside (or on) the unit circle.  Hence the pencil beam design shown in Figure \ref{pencil_FIR} can be classified as a {minimum phase} linear array\footnote{The Taylor pattern design proposed by Villeneuve \cite{villeneuve} also belongs to the class of minimum phase linear arrays, as all zeros are located on the unit circle. }.  The class of minimum phase arrays has a number of useful properties, of which the most important one (for array design) is that it contains the minimal number of array elements to achieve the given array radiation magnitude specifications \cite{oppenheim,oppenheim2,linear_syst}.  No other array can both satisfy the specifications and deploy  a smaller number of array elements. 

To make this statement precise,  let there be $S$ element excitation vectors  $\mathbf c_{q \in \{1,\cdots,S\}}$ with identical magnitude radiation patterns.  Then there is a unique minimum phase vector, denoted $\mathbf c_p$, which has the least number of elements --- in the sense that for any $N$ element coefficients and  $0 \le k \le N-1$, then it can be shown that \cite{smith}
\begin{equation}
\sum_{n=0}^{k} \, |{c_p[n]}|^2  >  \sum_{n=0}^{k} \, |{c_q[n]}|^2 ~\forall~ q \neq p.
\end{equation}

This paper is concerned with the design of minimum phase linear arrays that have a flat top  \cite{flattop1,flattop2,flattop3}, which from a FIR filter design point of view implies the passband is finite.   The minimum phase FIR design developed in \cite{olivier} is able to yield a design for this case as well, which will be demonstrated in this paper.  The minimum phase design \cite{olivier} is based on matrix factorization, and requires that the sidelobe specification be symmetric in the $u$ domain.

The paper is organised as follows.  Section \ref{theory} provides the theoretical results and algorithm for the minimum phase design procedure. Section \ref{results} presents three designs with a comparison to results from the literature, and conclusions are presented in Section \ref{conclude}.  

\section{Minimum phase linear array design} \label{theory}

\begin{figure} []
\centering
  \includegraphics[width=0.5\textwidth]{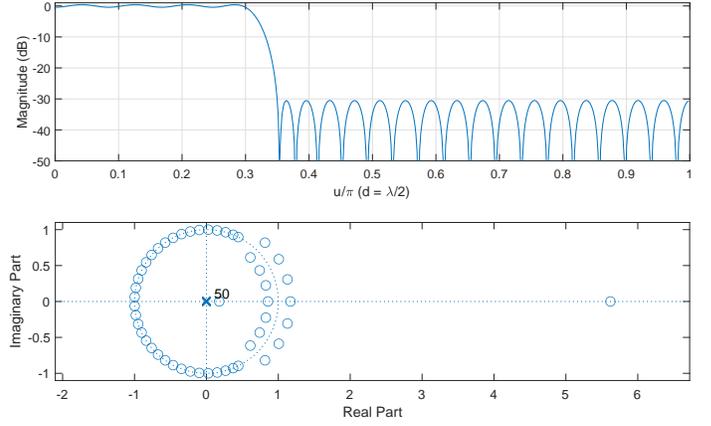}
\caption{A flat top linear phase array designed with the PM algorithm, with a constant sidelobe level specified at $-30$ dB. Clearly the flat top approximation is achieved by having reciprocal zeros inside and outside the unit circle.}
\label{flattop_example}
\end{figure}

Denote the radiation pattern of the desired  array as $C(\theta)$, and the corresponding minimum phase element coefficients by the vector $\mathbf c$.  The minimum phase design starts by performing a linear phase design with a radiation pattern given by $G(\theta)$, with a specification derived\footnote{Computing the specification for the linear phase array $G(\theta)$ given specifications for $C(\theta)$, and designing the linear phase array based on the PM algorithm is presented in detail in \cite{antonio}.} from $|C(\theta)|^2$ \cite{antonio}.  Any appropriate method can be used to design the linear phase array with radiation pattern $G(\theta)$, but in this paper the PM method is used which yields a Chebyshev approximation. An example flat top linear phase array designed based on the PM method is shown in Figure \ref{flattop_example}, where only the positive part of the $u$ range is shown.  
The flat top radiation pattern with linear phase is achieved by placing zeros outside the unit circle, as well as reciprocal zeros inside the unit circle.  

To design a minimum phase array, the linear phase array $G(\theta)$ with a $(2N -1)$ element excitation vector $\mathbf g$ is factorized, which yields $\mathbf c$ with $N$ elements. All zeros of $\mathbf c$ are inside (or on) the unit circle. The linear phase vector $\mathbf g$ is symmetric with the maximum amplitude element located at  $g[N-1]$. 

It was shown in \cite{olivier} that the element settings $\mathbf c$ for the minimum phase array can be obtained through Cholesky factorization of an Hermitian matrix $\mathbf G$, which represents $\mathbf g$.  The matrix is  given by\footnote{The vector $\mathbf g$ is real and symmetric.} 
\begin{eqnarray} \small  
\mathbf G = \left[ \begin{array}{lllllllllll}
g_{N-1}  & g_{N-2} & \cdots & g_{0} & 0 & \cdots & 0  \\
g_{N-2} & g_{N-1} & g_{N-2} & \cdots & g_{0} &  \cdots & 0 \\
\vdots & \vdots  & \vdots & \vdots &  \vdots  & \vdots  & \vdots  \\
g_0 & g_1 & \cdots  & g_{N-1} &  g_{N-2} & \cdots & \cdots  \\
0 & g_0 & \cdots & g_{N-2} & g_{N-1}  & g_{N-2}  & \cdots    \\
\vdots & \vdots  & \vdots & \vdots &  \vdots  & \vdots  & \vdots  \\
0  & \cdots & 0  & g_0 & \cdots &  g_{N-2} & g_{N-1} \\
\end{array} \right]
 \end{eqnarray}
and has $2Q + N$ rows.  $Q$ denotes the so-called expansion factor \cite{olivier}, required to achieve symmetry point equilibrium for the Cholesky factor matrix $\mathbf C$ obtained from factorising
\begin{equation}
    \mathbf C^\dagger \mathbf C = \mathbf G + \gamma \mathbf I.
\end{equation}
The parameter $\gamma$ is defined below.  It was demonstrated in \cite{olivier} that Cholesky factorization, in the limit where $Q \rightarrow {\infty}$, solves the non-linear equations given by 
\begin{eqnarray} \label{nonlinear_system_again}
\nonumber c_0^2 + \cdots + c_{N-1}^2 &=& g\left [{N-1}\right ] + \gamma \\ 
\nonumber c_0 c_1 + \cdots + c_{N-2} c_{N-1}  &=&  g\left [{N-2}\right ] \\
c_0 c_2 + \cdots + c_{N-3} c_{N-1} &=&  g\left [{N-3}\right ] \\
\nonumber \vdots  &=&  \vdots \\
\nonumber c_0 c_{M-1} &=& g[0]. 
\end{eqnarray}
The parameter $\gamma$ must be chosen as \cite{olivier}
\begin{equation}
    \gamma > |\lambda_{\mathrm{min}}|
\end{equation}
where $\lambda_{\mathrm{min}}$ represents the most negative eigenvalue of $\mathbf G$ -- if the smallest eigenvalue is positive (which is rarely the case) then $\gamma = 0$.  It was shown in \cite{olivier} that a numerical solution of the equations given by (\ref{nonlinear_system_again}) yields a very accurate design, with residual errors at the lower limit set by the digital machine resolution.  

In practice accuracy to $8$ decimal places for the weights $\mathbf c$ that are obtained through the process described above may not be required, and under those conditions a direct  approach can be used instead \cite{olivier}.  To obtain $\mathbf c$ directly, first define an augmented Kronecker delta given by 
\begin{equation}
\mathbf \delta_\mathrm{aug} = \{\underbrace{0,\cdots,0}_{\text{$Q$ zeros}}, {\underbrace{0,\cdots,0}_{\text{${N-1}$ zeros}}},1,\underbrace{0,\cdots,0}_{\text{$Q$ zeros}} \}^{\mathrm T}.
\end{equation}
The augmented form of the minimum phase linear array element weights is then given by 
\begin{equation}
\mathbf c_\mathrm{aug} = \{\underbrace{0,\cdots,0}_{\text{$Q$ zeros}},c_{M-1},\cdots, c_1,c_0,\underbrace{0,\cdots,0}_{\text{$Q$ zeros}} \}^{\mathrm T}.
\end{equation}
For a sufficiently large value of $Q$, it follows that  
 \cite{olivier}
\begin{equation} \label{transf}
\mathbf c_{\mathrm{aug}}  =  \mathbf C~ \mathbf \delta_{\mathrm{aug}}. % 
\end{equation}
By purging the $2Q$ zeros the minimum phase weights $\mathbf c$ can be obtained. 

In this paper the procedure presented above will be used in the three example designs presented in Section \ref{results},   and in all cases $Q$ was set to $Q = 30 N$.

\section{Experimental results for minimum phase array design} \label{results}

\subsection{Design $1$}

This section presents the design of a minimum phase linear array for the following specifications obtained from an example design in \cite{flattop2}:

\begin{enumerate}
    \item Element spacing at half wavelength.
    \item A flat-top ripple less than $0.25$ dB.
    \item Sidelobes may not exceed $-52$ dB.
    \item A main beam with steering direction at $\theta = 0$. 
    \item Flat-top edge located at  $u=\pi \sin(0.2182)$.
    \item The first main beam zero at $u =\pi \sin(\frac{\pi}{3})$.
\end{enumerate}

\begin{figure} []
\centering
  \includegraphics[width=0.4\textwidth]{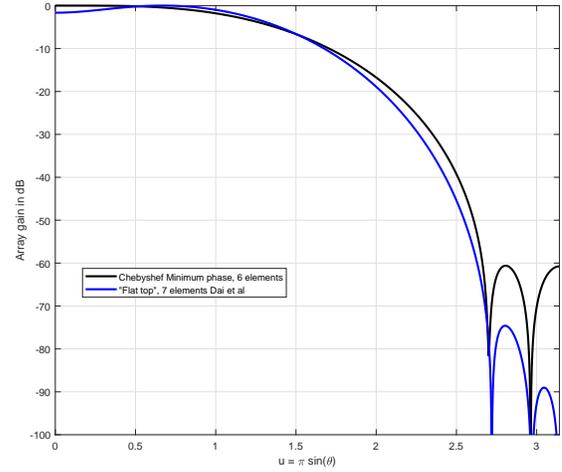}
\caption{A flat top minimum phase array designed to the given specifications. The results show that the minimum phase linear array exceeds the specifications with $6$ array elements.  The result is compared to the design in \cite{flattop2} where $7$ array elements were deployed. }
\label{dai}
\end{figure}

In \cite{flattop2} the authors restricted the design to $N=7$ elements, however for the minimum phase linear array the required order (number of elements) is obtained as part of the design procedure  \cite{olivier}.  The minimum phase array weights $\mathbf c$ obtained as proposed in Section \ref{theory} yielded $6$ array elements, and the radiation pattern so obtained exceeds the given specifications, as shown in Figure \ref{dai}. Note that the flat top ripple for the minimum phase array is significantly lower than that obtained in \cite{flattop2}, and the sidelobes remain below the specified bound.

The design proposed in \cite{flattop2} combines the  Dolph-Chebyshev zeros on the unit circle with zeros not located on the unit circle, which then realizes a flat-top main beam shape.   The comparison with a minimum phase linear array shown in Figure \ref{dai}  clearly shows that the minimum phase linear array requires fewer array elements to achieve the specifications. 

Note that the specifications provided in \cite{flattop2} did not require the sidelobes to taper off, and hence a Chebyshev design was deployed.  If the sidelobes are required to taper off, then the minimum phase Chebychev element weights $\mathbf c$ can be optimized to have sidelobes that taper off \cite{olivier2}. 

\subsection{Design $2$}

In this section the linear phase array design proposed in \cite{flattop3} is compared to a minimum phase array design, which will demonstrate the substantial gains that can be achieved if the phase is not constrained to be linear. The comparison will thus be between a linear phase design presented in \cite{flattop3} and the proposed minimum phase array design based on the same specifications. Note that once again the sidelobes were not specified to taper off \cite{flattop3}, and thus a Chebyshev minimum phase design will be performed as proposed in Section \ref{theory}.  

\begin{figure} []
\centering
  \includegraphics[width=0.38\textwidth]{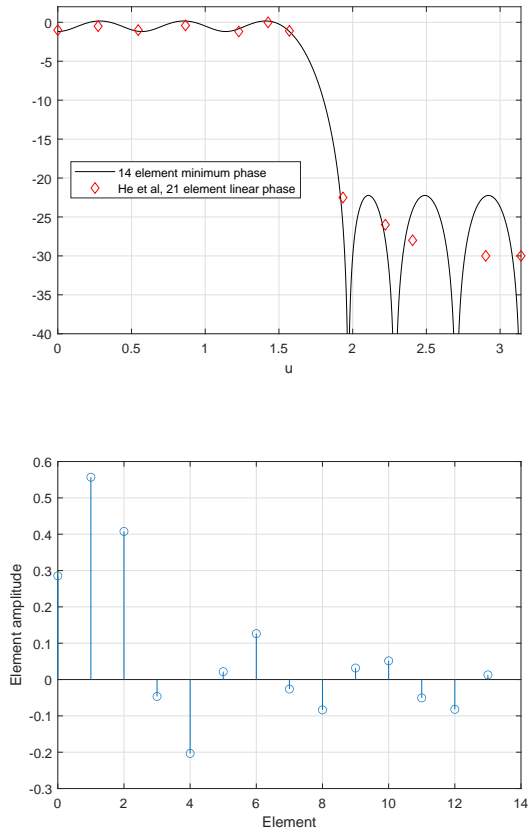}
\caption{A flat-top linear array with a maximum sidelobe specification of $-21$ dB and a maximum flat top ripple of $1.18$ dB, from \cite{flattop3}. The linear phase array design presented in \cite{flattop3} required 21 elements, while a minimum phase array requires only $14$ elements. }
\label{he}
\end{figure}

The specifications provided are as follows \cite{flattop3}:
\begin{enumerate}
    \item Element spacing at half wavelength.
    \item A flat-top ripple less than $1.18$ dB.
    \item Sidelobes may not exceed $-21$ dB for $u \ge 1.92$.
    \item A main beam with steering direction at $\theta = 0$. 
    \item Flat-top width equal to $60$ degrees.
\end{enumerate}
The magnitude of the radiation pattern is shown in Figure \ref{he}, along with radiation magnitude values from \cite{flattop3}.  Both designs exceed the specifications, but it should be obvious that the minimum phase array required only $14$ elements, while the linear phase array \cite{flattop3} required $21$ elements. The element excitation amplitudes are also shown in Figure \ref{he}.  As it is a minimum phase array, the weights tend to have energy concentrated in a few elements, which is typical for this class of array. 

\subsection{Design 3}

In this  subsection a non-symmetric design from \cite{orchard} is considered, with specifications given as follows:
\begin{enumerate}
    \item Element spacing at half wavelength.
    \item Flat-top width $25$ degrees.
    \item A flat-top ripple less than $0.5$ dB.
    \item Sidelobes may not exceed $-30$ dB for negative angle, and not exceed $-20$ dB for positive angles. 
    \item A main beam with steering direction at $\theta = 0$.
\end{enumerate}

\begin{figure} []
\centering
  \includegraphics[width=0.5\textwidth]{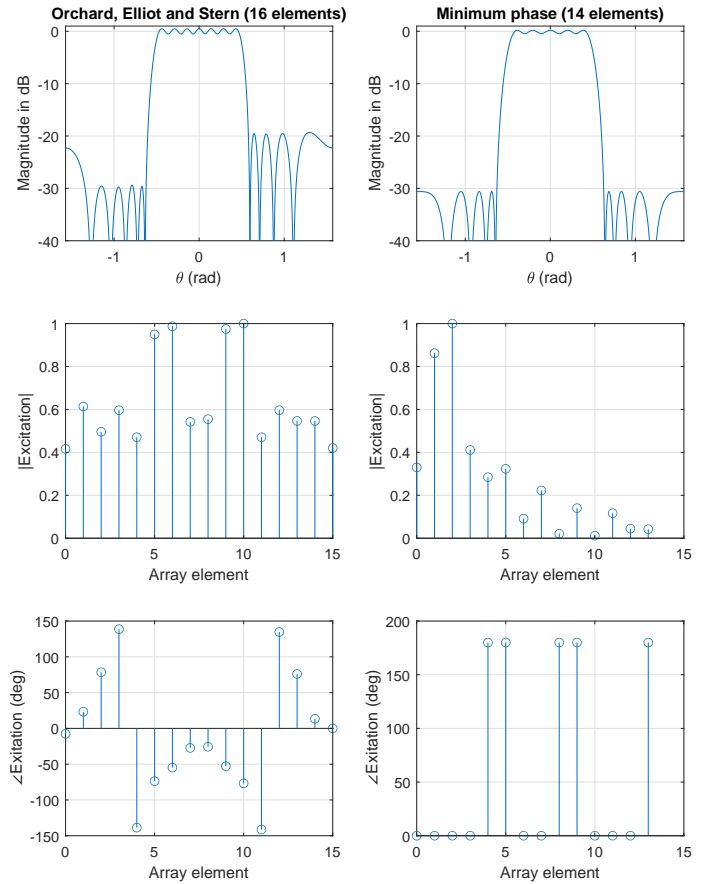}
\caption{A flat-top linear array with uneven sidelobes, from an array design in \cite{orchard}. Note that the phases of the minimum phase array weights are either $0$ or $\pi$.  The weight magnitude shown was normalized relative to the maximum weight.}
\label{transform}
\end{figure}

In \cite{orchard} a $16$ element array with complex element weights was designed to meet this specification, and is shown on the left in Figure \ref{transform}.  A minimum phase array with real element weights was designed able to exceed the specifications as shown in Figure \ref{transform}, and required only $14$ elements. If for some reason the array radiation pattern is required to have higher sidelobes for positive angles, then additional optimization yielding complex weights  can be performed \cite{olivier2}.  

%Note the phase of the elements for the minimum phase array, which are all at either $0$ or $180$ degrees -- this may aid realization in the presence of mutual coupling.

\section{Conclusions} \label{conclude}

Based on recent advances in the design of a minimum phase Finite Impulse Response (FIR) filter \cite{olivier}, the class of minimum phase linear arrays and its design were presented and studied.  A straightforward procedure to design these arrays is provided in this paper, based on Cholesky factorization of a Hermitian matrix representing a linear phase array.  Three example designs each with a flat top were presented and compared to results from the literature.  This demonstrated that as expected, the minimum phase linear array requires the least number of array elements of all linear arrays able to achieve the magnitude radiation pattern specification.

\bibliography{mybibfile}

\end{document}